\documentclass[aps,prb,superscriptaddress,showpacs,floatfix,twocolumn]{revtex4-1}
\usepackage{graphicx,amsmath,amssymb,xspace,epsfig,float,multirow,subfigure,tabularx}

\pdfoutput=1

\begin{document}

\title{Quantum critical point in the superconducting transition on the surface
of topological insulator }
\author{Dingping Li}
\email{lidp@pku.edu.cn}
\affiliation{School of Physics, Peking University, Beijing
100871, \textit{China}} \affiliation{Collaborative Innovation Center of
Quantum Matter, Beijing, China}
\author{Baruch Rosenstein}
\email{vortexbar@yahoo.com,correspondent author}
\affiliation{Electrophysics Department, National Chiao Tung University, Hsinchu 30050, \textit{Taiwan,
R. O. C}} \affiliation{Physics Department, Ariel University, Ariel 40700, Israel}
\author{I. Shapiro}
\affiliation{Physics Department, Bar-Ilan University, 52900 Ramat-Gan, Israel}
\author{B.Ya. Shapiro}
\email{shapib@mail.biu.ac.il}
\affiliation{Physics Department, Bar-Ilan University, 52900 Ramat-Gan, Israel}

\begin{abstract}
Pairing in the Weyl semi - metal appearing on the surface of topological
insulator is considered. It is shown that due to an "ultra-relativistic"
dispersion relation there is a quantum critical point governing the zero
temperature transition to a superconducting state. Starting from the
microscopic Hamiltonian with local attraction, we calculated using the
Gor'kov equations, the phase diagram of the superconducting transition at
arbitrary chemical potential, its magnetic properties and critical exponents
close to the quantum critical point. The Ginzburg - Landau effective theory
is derived for small chemical potential allowing to consider effects of
spatial dependence of order parameters in magnetic field. The GL equations
are very different from the conventional ones reflecting the chiral
universality class of the quantum phase transition. The order parameter
distribution of a single vortex is found to be different as well. The
magnetization near the upper critical field is found to be quadratic, not
linear as usual. We discuss the application of these results to recent
experiments in which surface superconductivity was found that some 3D
topological insulators and estimate feasibility of the phonon pairing.
\end{abstract}

\pacs{74.20.Fg, 74.90.+n, 74.20.Op }
\maketitle

%%\author{D. P. Li$^{1},$ B., B. Rosenstein$^{2,3}$, I. Shapiro$^{4}$ and Ya.
%Shapiro$^{4\ast }$ }

\section{Introduction}

Topological insulator (TI) is a novel state of matter in materials with
strong spin - orbit interactions that create topologically protected surface
states \cite{Zhang}. The electrons (holes) in these states have a linear
dispersion relation and can be described approximately by a (pseudo)
relativistic two dimensional (2D) Weyl Hamiltonian. The system with the
chemical potential above or below the Weyl point realizes an
"ultra-relativistic" 2D electron or hole conducting liquid. It is known for
a long time that similar 2D and quasi-2D metallic systems like the surface
metal on twin planes \cite{Shapiro}, layered materials (strongly anisotropic
high $T_{c}$ cuprates\cite{Wen} or organic superconductors\cite{organic})
may develop 2D (surface) superconductivity. This phenomenon became known as
"localized superconductivity"\cite{Buzdin}. Since best studied TIs possess a
quite standard phonon spectrum \cite{phononexp}, it was predicted recently
\cite{DasSarma} that they become superconducting TI (STI) (this should be
distinguished from "topological superconductors", TSC, in which
superconductivity appears in the bulk \cite{Zhang}). The predicted critical
temperature of order of $1K$ is rather low (despite a fortunate suppression
of the Coulomb repulsion due to a large dielectric constant $\varepsilon
\sim 50$), the nature of the "normal" state (so-called 2D Weyl semi-metal)
might make the superconducting properties of the system unusual. The
ultra-relativistic nature manifests itself mostly when the Weyl cone is very
close to the Fermi surface. Especially interesting is the case (that
actually was originally predicted for the [111] surface of $Bi_{2}Te_{3}$
and $Bi_{2}Se_{3}$\cite{Zhang1}) when the chemical potential coincides with
the Weyl point. Although subsequent ARPES experiments\cite{Zhang} show the
location of the cone of surface states order tenths of $eV$ off the Fermi
surface; there are experimental means to shift the chemical potential, for
example by the bias voltage \cite{bias}.

Unlike the more customary poor 2D metals with several small pockets of
electrons/holes on the Fermi surface (in semiconductor systems or even some
high $T_{c}$ materials\cite{Wen}), the electron gas STI has two
peculiarities especially important when pairing is contemplated. The first
is the bipolar nature of the Weyl spectrum: there is no energy gap between
the upper and lower cones. The second is that \ the spin degree of freedom
is a major player in the quasiparticle dynamics. This degree of freedom
determines the pairing channel. The pairing channel problem was studied
theoretically on the level of the Bogoliubov-deGennes equation \cite{Herbut}%
. Both $s$-wave and $p$-wave are possible and compete due to the breaking of
the bulk inversion symmetry by the surface. The spectrum of Andreev states
of the Abrikosov vortex was obtained\cite{Nori} in a related problem of TI
in contact with an $s$-wave superconductor \cite{DasSarma1}. Various pairing
interactions were considered to calculate the DOS measured in $%
Cu_{x}Bi_{2}Se_{3}$ to discriminate between STI and TSC using
self-consistent analysis \cite{Sato}. As mentioned above the most intriguing
case is that of the small chemical potential that has not been addressed
microscopically. It turns out that it is governed by{\LARGE \ }a quantum
critical point (QCP)\cite{Sachdev}.

The concept of QCP at zero temperature and varying doping constitutes a very
useful language for describing the microscopic origin of superconductivity
in high $T_{c}$ cuprates and other "unconventional" superconductors\cite{Wen}%
. Superconducting transitions generally belong to the $U\left( 1\right) $
class of second order phase transitions\cite{Landau}, however it was pointed
out a long time ago\cite{Rosenstein} that, if the normal state dispersion
relation is "ultra-relativistic", the transition at zero temperature as
function of parameters like the pairing interaction strength is
qualitatively distinct and belongs to chiral universality classes classified
in ref. \cite{Gat}. Attempts to experimentally identify second order
transitions governed by QCP included quantum magnets \cite{Sachdev},
superconductor - insulator transitions\cite{SCinsulator} and more recently
chiral condensate in graphene\cite{Katsnelson,CastroNeto}.

In this paper we study the thermodynamic and magnetic properties of the
surface superconductivity in TI with local attraction pairing Hamiltonian
characterized by the coupling strength $g$ and cutoff parameter $T_{D}$
within the self-consistent approximation. The phase diagram for $s$-wave
pairing is obtained for arbitrary temperature $T$ and chemical potential $%
\mu <T_{D}$. The latter condition is the main difference from the
conventional BCS model in which $\mu >>T_{D}$. We found a quantum critical
point at $T=\mu =0$ when the coupling strength $g$ reaches a critical value $%
g_{c}$ dependent on the cutoff parameter. We concentrate on properties of
the superconducting state in a part of the phase diagram that is dominated
by the QCP. Various critical exponents are obtained. In particular, the
coupling strength dependence of the coherence length is $\xi \propto \left(
g-g_{c}\right) ^{-\nu }$ with $\nu =1$ , the order parameter scales as $%
\Delta \propto \left( g-g_{c}\right) ^{\beta }$, $\beta =1$. It is found
that near the QCP the Ginzburg - Landau effective model is rather
unconventional. The structure of the single vortex core is different from
the usual Abrikosov vortex, while the magnetization curve near the upper
critical magnetic field $H_{c2}$ is quadratic: $M=\left( H-H_{c2}\right)
^{2} $, not linear.

The rest of the paper is organized as follows. The model and the method of
its solution (in the Gorkov equations form) are presented in Section II. The
phase diagram in the homogeneous case (no magnetic field) is established and
the unusual nature of the phase transition discussed. The novel case of zero
chemical potential (tuning to the Weyl point) is studied in detail. The
Ginzburg-Landau energy is derived in Section III and exploited to determine
magnetic properties of STI. The $H_{c2}$ line and magnetization curves for a
dense vortex lattice as well as the single vortex texture are obtained.
Section IV contains discussion on experimental feasibility of the phonon
mediated surface superconductivity in TI, comparison with more familiar BEC
and BCS scenarios and\ conclusion.

\section{The s-wave pairing model. The phase diagram.}

\subsection{TI in magnetic field with a local pairing interaction. Gor'kov
equations.}

Electrons on the surface of a TI perpendicular to $z$ axis, see Fig.1, are
described by a Pauli spinor $\psi _{\alpha }\left( \mathbf{r}\right) $,
where the upper plane, $\mathbf{r}=\left\{ x,y\right\} $, is considered,
with spin projections taking the values $\alpha =\uparrow ,\downarrow $ with
respect to $z$ axis. The Hamiltonian for electrons in TI subjected to a
perpendicular external homogeneous magnetic field, and interacting via
four-Fermi local coupling of strength $g$ is
\begin{eqnarray}
H &=&\int d^{2}r\left\{ \psi _{\alpha }^{+}\left( \mathbf{r}\right) \widehat{%
H}_{\alpha \beta }\psi _{\beta }\left( \mathbf{r}\right) \right.
\label{Hamiltonian} \\
&&\left. \mathbf{-}\frac{g}{2}\psi _{\alpha }^{+}\left( \mathbf{r}\right)
\psi _{\beta }^{+}\left( \mathbf{r}\right) \psi _{\beta }\left( \mathbf{r}%
\right) \psi _{\alpha }\left( \mathbf{r}\right) \right\} +H_{mag}\text{.}
\notag
\end{eqnarray}
\begin{figure}[tbp]
\centering
\includegraphics[width=8cm]{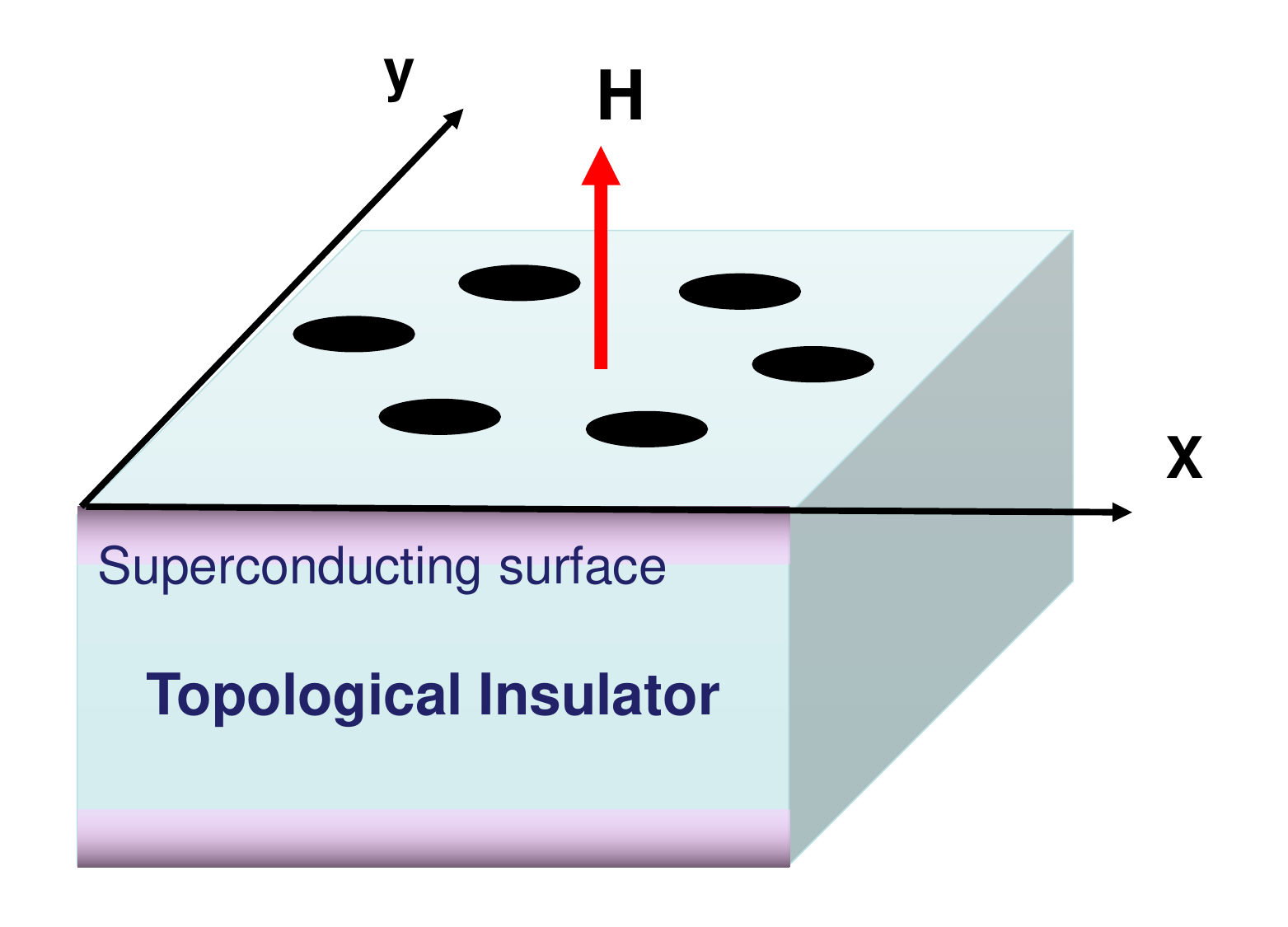} \vspace{-0.5cm}
\caption{Topological insulator plate in magnetic field. Surfaces are
populated by Weyl quasiparticles and holes that both can be paired by
interactions. Magnetic field creates vortices with normal cores (dark areas
on the surfaces) of the radius of order of coherence length $\protect\xi $.}
\end{figure}
Here the surface Weyl Hamiltonian matrix \cite{Zhang,Herbut} is defined as%
\begin{eqnarray}
\widehat{H}_{\alpha \beta } &=&v_{F}\varepsilon _{ij}\widehat{P}_{i}\sigma
_{\alpha \beta }^{j}-\mu \delta _{\alpha \beta }\text{;}  \label{Weyl} \\
\widehat{\mathbf{P}} &\equiv &-i\hbar \mathbf{\nabla -}\frac{e^{\ast }}{c}%
\mathbf{A}\left( \mathbf{r}\right) \text{,}  \notag
\end{eqnarray}%
where $i,j=x,y$; $v_{F}$ is the Fermi velocity of the TI and $\mu $ is the
surface chemical potential. $\sigma ^{j}$ are the Pauli matrices and $%
\varepsilon _{ij}$ is the antisymmetric tensor. Only one valley is
explicitly considered (generalization to several "flavors" is trivial).
Vector potential $\mathbf{A}$ describes the 3D magnetic induction $\mathbf{B}%
=\mathbf{\nabla \times A}$ with magnetic energy given by%
\begin{equation}
H_{mag}=\frac{1}{8\pi }\int d^{2}rdz\left( \mathbf{B}\left( \mathbf{r,}%
z\right) -\mathbf{H}_{ext}\right) ^{2}\text{.}  \label{Hmag}
\end{equation}

The effective local interaction might be generated by a phonon exchange or
perhaps other mechanisms and will be assumed to be weak coupling. Therefore
the BCS type approximation can be employed. Using the standard formalism,
the Matsubara Green's functions ($\tau $ is the Matsubara time),
\begin{eqnarray}
G_{\alpha \beta }\left( \mathbf{r},\tau ;\mathbf{r}^{\prime },\tau ^{\prime
}\right)  &=&-\left\langle T_{\tau }\psi _{\alpha }\left( \mathbf{r},\tau
\right) \psi _{\beta }^{\dagger }\left( \mathbf{r}^{\prime },\tau ^{\prime
}\right) \right\rangle \text{;}  \label{GFdef} \\
F_{\alpha \beta }^{\dagger }\left( \mathbf{r},\tau ;\mathbf{r}^{\prime
},\tau ^{\prime }\right)  &=&\left\langle T_{\tau }\psi _{\alpha }^{\dagger
}\left( \mathbf{r},\tau \right) \psi _{\beta }^{\dagger }\left( \mathbf{r}%
^{\prime },\tau ^{\prime }\right) \right\rangle \text{,}  \notag
\end{eqnarray}%
obey the Gor'kov equations\cite{AGD}:%
\begin{gather}
-\frac{\partial G_{\gamma \kappa }\left( \mathbf{r},\tau ;\mathbf{r}^{\prime
},\tau ^{\prime }\right) }{\partial \tau }-\int_{\mathbf{r}^{\prime \prime
}}\left\langle \mathbf{r}\left\vert \widehat{H}_{\gamma \beta }\right\vert
\mathbf{r}^{\prime \prime }\right\rangle G_{\beta \kappa }\left( \mathbf{r}%
^{\prime \prime },\tau ;\mathbf{r}^{\prime },\tau ^{\prime }\right)
\label{Gorkov} \\
-gF_{\beta \gamma }\left( \mathbf{r},\tau ;\mathbf{r},\tau \right) F_{\beta
\kappa }^{\dagger }\left( \mathbf{r},\tau ,\mathbf{r}^{\prime },\tau
^{\prime }\right) =\delta ^{\gamma \kappa }\delta \left( \mathbf{r-r}%
^{\prime }\right) \delta \left( \tau -\tau ^{\prime }\right) ;  \notag \\
\frac{\partial F_{\gamma \kappa }^{\dagger }\left( \mathbf{r},\tau ;\mathbf{r%
}^{\prime },\tau ^{\prime }\right) }{\partial \tau }-\int_{\mathbf{r}%
^{\prime \prime }}\left\langle \mathbf{r}\left\vert \widehat{H}_{\gamma
\beta }^{t}\right\vert \mathbf{r}^{\prime \prime }\right\rangle F_{\beta
\kappa }^{\dagger }\left( \mathbf{r}^{\prime \prime },\tau ;\mathbf{r}%
^{\prime },\tau ^{\prime }\right)   \notag \\
-gF_{\gamma \beta }^{\dagger }\left( \mathbf{r},\tau ;\mathbf{r},\tau
\right) G_{\beta \kappa }\left( \mathbf{r},\tau ,\mathbf{r}^{\prime },\tau
^{\prime }\right) =0\text{.}  \notag
\end{gather}%
In the presence of magnetic field these equations are complicated by
emergence of inhomogeneity pertinent to type II superconductors. This will
be addressed in Section III. Here we solve the homogeneous case when no
magnetic field is present.

\subsection{Uniform condensate.}

In the homogeneous case the Gor'kov equations for Fourier components of the
Greens functions simplify considerably,
\begin{eqnarray}
D_{\gamma \beta }^{-1}G_{\beta \kappa }\left( \omega ,p\right) -\widehat{%
\Delta }_{\gamma \beta }F_{\beta \kappa }^{\dagger }\left( \omega ,p\right)
&=&\delta ^{\gamma \kappa }\text{;}  \label{Gorkov_uniform} \\
D_{\beta \gamma }^{-1}F_{\beta \kappa }^{\dagger }\left( \omega ,p\right) +%
\widehat{\Delta }_{\gamma \beta }^{\ast }G_{\beta \kappa }\left( \omega
,p\right)  &=&0\text{,}  \notag
\end{eqnarray}%
where $\omega =\pi T\left( 2n+1\right) $ is the Matsubara frequency and$\
D_{\gamma \beta }^{-1}=\left( i\omega -\mu \right) \delta _{\gamma \beta
}-v_{F}\varepsilon _{ij}p_{i}\sigma _{\alpha \beta }^{j}$. The matrix gap
function can be chosen as ($\Delta $ real)
\begin{equation}
\widehat{\Delta }_{\beta \gamma }=gF_{\gamma \beta }\left( 0\right) =\left(
\begin{array}{cc}
0 & \Delta  \\
-\Delta  & 0%
\end{array}%
\right) \text{.}  \label{delta}
\end{equation}

These equations are conveniently presented in matrix form (superscript $t$
denotes transposed and $I$ - the identity matrix):
\begin{eqnarray}
D^{-1}G-\widehat{\Delta }F^{\dagger } &=&I\text{;}  \label{matrixeq} \\
D^{t-1}F^{\dagger }+\widehat{\Delta }^{\ast }G &=&0\text{.}  \notag
\end{eqnarray}%
Solving these equations one obtains
\begin{eqnarray}
G^{-1} &=&D^{-1}+\widehat{\Delta }D^{t}\widehat{\Delta }^{\ast }\text{;}
\label{solution} \\
F^{\dagger } &=&-D^{t}\widehat{\Delta }^{\ast }G\text{,}  \notag
\end{eqnarray}%
with the gap function found from the consistency condition
\begin{equation}
\widehat{\Delta }^{\ast }=-g\sum\limits_{\omega q}D^{t}\widehat{\Delta }%
^{\ast }G\text{.}  \label{gap eq}
\end{equation}%
The off-diagonal component of this equation is:
\begin{align}
\Delta & =g\Delta \sum\limits_{\omega q}\left( \Delta
^{2}+v_{F}^{2}p^{2}+\mu ^{2}+\hbar ^{2}\omega ^{2}\right)   \label{gap} \\
& \times \frac{1}{\left( \Delta ^{2}+\hbar ^{2}\omega ^{2}+\left( v_{F}p-\mu
\right) ^{2}\right) \left( \Delta ^{2}+\hbar ^{2}\omega ^{2}+\left(
v_{F}p+\mu \right) ^{2}\right) }\text{.}  \notag
\end{align}

The spectrum of elementary excitations obtained from the poles of the Greens
function coincides with that found within the Bogoliubov - de Gennes
approach \cite{Herbut}
\begin{equation}
E_{p}=\pm \sqrt{\Delta ^{2}+\left( v_{F}p-\mu \right) ^{2}}\text{.}
\label{spectrum}
\end{equation}%
The solutions of the gap equation are presented in the next subsection for a
general chemical potential and zero temperature, while more general
situations (arbitrary temperature and magnetic field) in the most
interesting case of $\mu =0$ are addressed in the next section.
\begin{figure}[tbp]
\centering
\includegraphics[width=8cm]{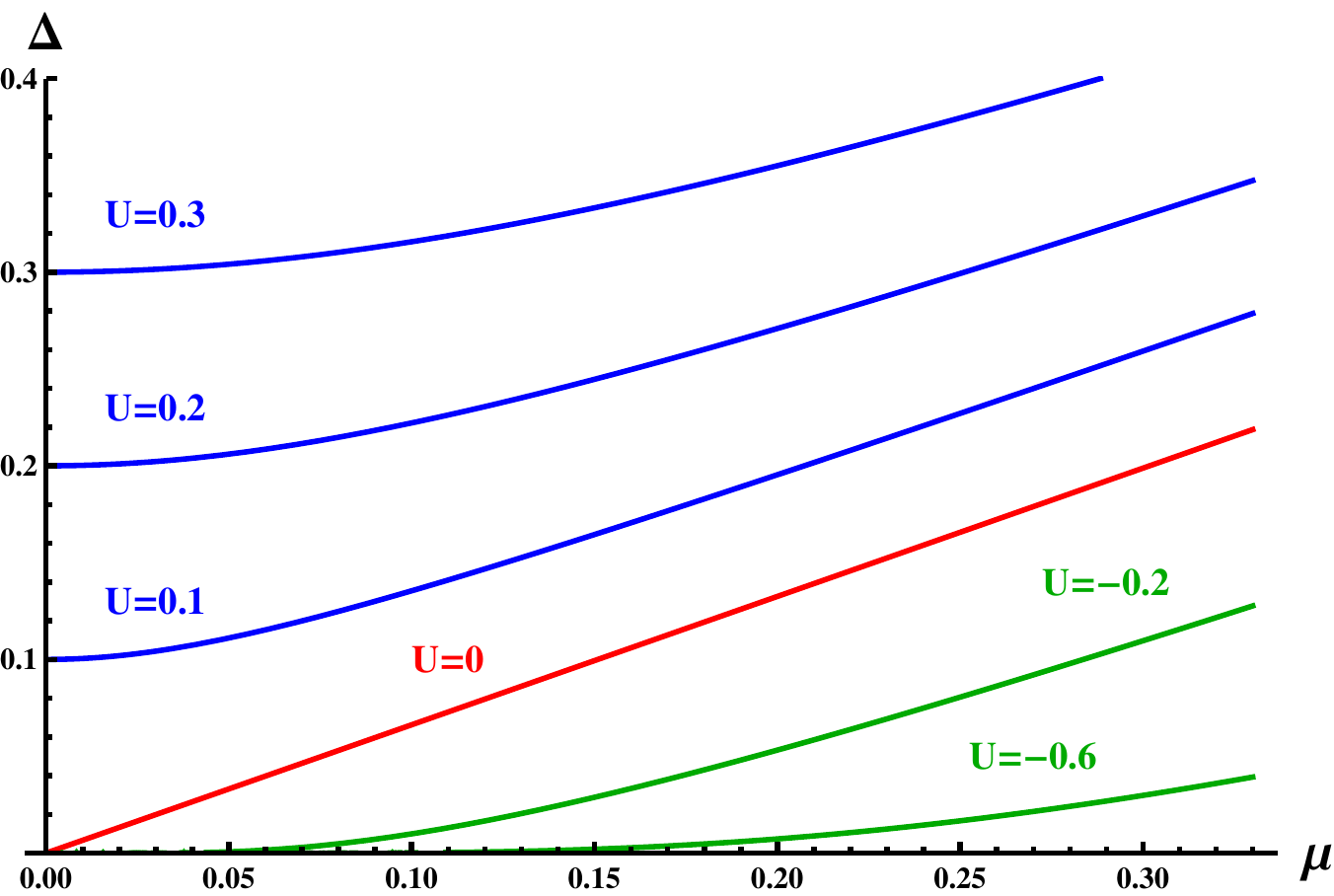}
\caption{Order parameter at zero temperature as function of chemical
potential of the TI surface Weyl semi-metal at various values of coupling
parametrized by the renormalized energy $U$, Eq.(\protect\ref{gren}). For
positive $U$ (blue lines) the superconductivity is strong and does not
vanish even for zero chemical potential. There exists the critical coupling,
$U=0$ (red line), at which the second order transition occurs at quantum
critical point $\protect\mu =0$. For negative $U$ the superconductivity
still exists at $\protect\mu >0$, but is exponentially weak. }
\end{figure}

\subsection{Zero temperature phase diagram and QCP.}

At zero temperature the integrations over frequency and momentum limited by
the UV cutoff $\Lambda $ result in (see Appendix A for details)
\begin{equation}
U\Delta =\Delta \left( \sqrt{\Delta ^{2}+\mu ^{2}}-\frac{\mu }{2}\log \frac{%
\sqrt{\Delta ^{2}+\mu ^{2}}+\mu }{\sqrt{\Delta ^{2}+\mu ^{2}}-\mu }\right)
\text{,}  \label{gapeq1}
\end{equation}%
where the dependence on the cutoff is incorporated in the renormalized
coupling with dimension of energy defined as
\begin{equation}
U=v_{F}\Lambda -\frac{4\pi \hbar ^{2}v_{F}^{2}}{g}\text{.}  \label{gren}
\end{equation}%
This can be interpreted as an effective binding energy of the Cooper pair in
the Weyl semi - metal. For concreteness we consider only $\mu >0$, although
the particle - hole symmetry makes the opposite case of the hole doping, $%
\mu <0$, identical. Of course the superconducting solution exists only for $%
g>0$. In Fig. 2 the dependence of the gap $\Delta $ as function of the
chemical potential $\mu $ is presented for different values of $U$.

For an attractive coupling $g$ stronger than the critical one,
\begin{equation}
g_{c}=\frac{4\pi \hbar ^{2}v_{F}}{\Lambda }\text{,}  \label{g_c}
\end{equation}%
(when $U>0$), blue lines in Fig. 2, there are two qualitatively different
cases.

(i). When $\mu <<U$ the dependence of $\Delta $ on the chemical potential is
parabolic, see Appendix B:%
\begin{equation}
\frac{\Delta }{U}\approx 1+\left( \frac{\mu }{U}\right) ^{2}\text{.}
\label{delta_neg}
\end{equation}

\begin{figure}[tbp]
\centering
\includegraphics[width=8cm]{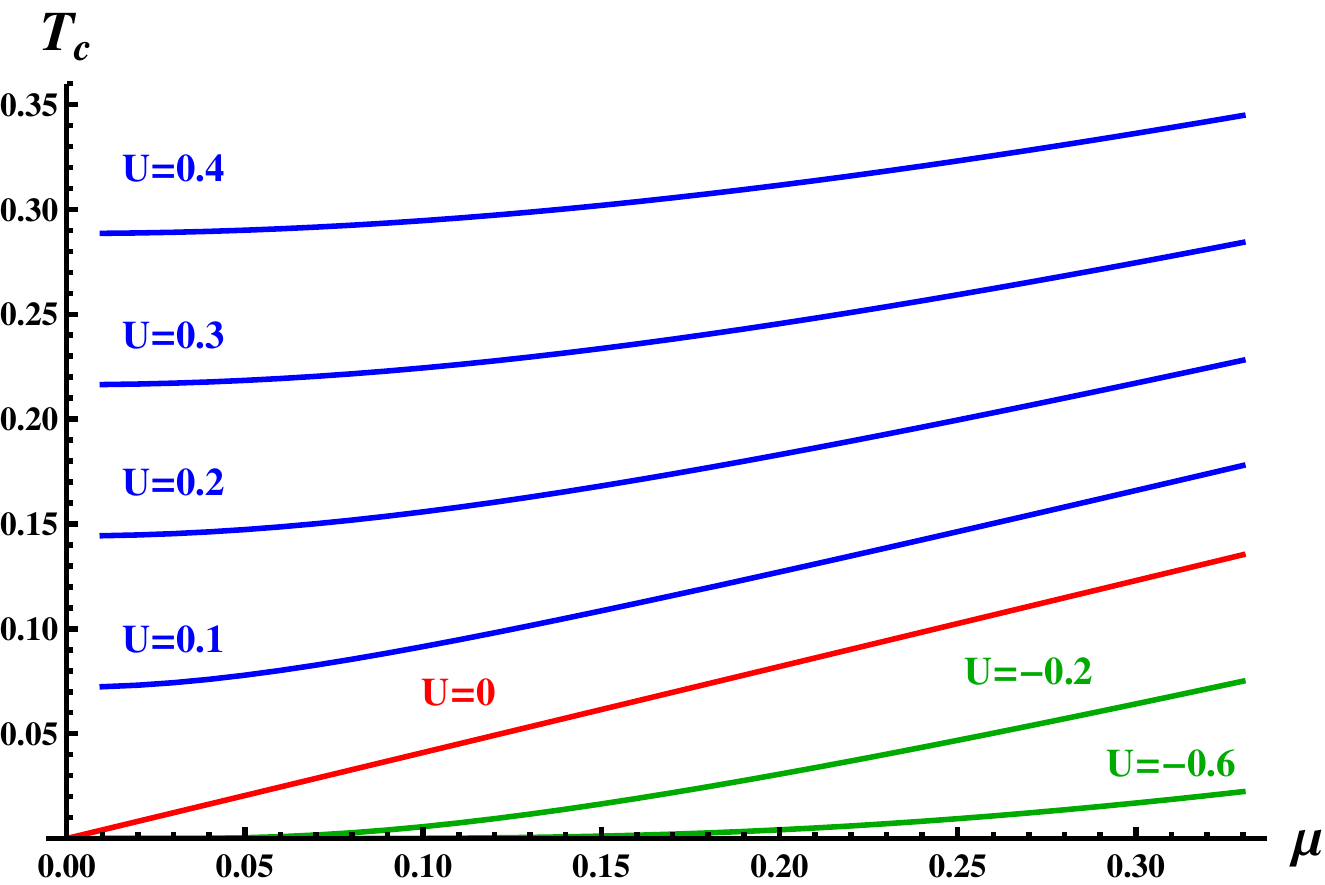}
\caption{Transition temperature as function of chemical potential at
supercritical ($U>0$, in blue), critical ($U=0$, in red) and subcritical
values of coupling. }
\end{figure}

In particular, when $\mu =0,$ the gap equals $U$. As can be seen from Fig.
2, the chemical potential makes a very limited impact in the large portion
of the phase diagram.

(ii) For the attraction just stronger than critical, $g>g_{c}$, namely for
small positive $U$, the dependence becomes linear, see red line in Fig. 2, $%
\Delta =0.663\,\mu $. So that the already weak condensate becomes sensitive
to $\mu $.

The case (i) is more interesting than (ii) since it exhibits stronger
superconductivity (larger $T_{c}$, see below). Finally for $g<g_{c}\,$,
namely negative $U$ (green lines in Fig. 2), the superconductivity is very
weak with exponential dependence similar to the BCS one,%
\begin{equation}
\Delta \approx \mu \text{ exp}\left[ -\left( \left\vert U\right\vert /\mu
-1\right) \right] \text{.}  \label{delta_pos}
\end{equation}%
More detailed comparisons will be performed in Section IV. As was mentioned
above, in the more interesting cases of large $\Delta $ the dependence on
the chemical potential is very weak. A peculiarity of superconductivity in
TI is that electrons (and holes) in Cooper pairs are created themselves by
the pairing interaction rather than being present in the sample as free
electrons. Therefore it is shown that it is possible to neglect the effect
of weak doping and consider directly the $\mu =0$ particle-hole symmetric
case. This point in parameter space is the QCP \cite{Sachdev} and will be
studied in detail in what follows. Of course, at finite temperature at any
attraction, $g>0$, there exists a (classical) superconducting critical point
at certain temperature $T_{c}$ that is calculated next.

\subsection{Dependence of the critical temperature $T_{c}$ on strength of
pairing interaction.}

Summation over Matsubara frequency and integrations over momenta in the gap
equation, Eq.(\ref{gapeq1}), at finite temperature and arbitrary chemical
potential are performed in Appendix B. The critical temperature as a
function of $\mu $ and (positive) $U$ is obtained numerically and presented
in Fig. 3. Again at relatively large $U$ the dependence of $T_{c}$ on the
chemical potential is very weak and parabolic. When $0<g<g_{c}$ the critical
temperature is exponentially small albeit nonzero.

\subsection{Zero chemical potential $\protect\mu =0$.}

At zero chemical potential the Hamiltonian Eq.(\ref{Hamiltonian}) possesses
a particle - hole symmetry. Microscopically, Cooper pairs of both electrons
and holes are formed. The system is unique in this sense since the electron
- hole symmetry is not spontaneously broken in both normal and
superconducting phases. Supercurrent in such a system does not carry
momentum or mass. Performing the sum and integral over momenta in the gap
equation, Eq.(\ref{gapeq1}), analytically (see Appendix A), it becomes
(using the definition of $U$ given in Eq.(\ref{gren})) for $U>0$:
\begin{equation}
U=2T\log \left[ 2\cosh \frac{\Delta }{2T}\right] \text{.}  \label{gapren_T}
\end{equation}%
At zero temperature $\Delta =U$, while $\Delta \rightarrow 0$ as a power of
the parameter $U\propto g-g_{c}$ describing the deviation from quantum
criticality
\begin{equation}
T_{c}=\frac{1}{2\log 2}U^{z\nu };\text{ \ }z\nu =1\text{.}  \label{Tc_mic}
\end{equation}%
Here $z$ is the dynamical critical exponent\cite{Sachdev}. Therefore, as
expected, the renormalized coupling describing the deviation from the QCP is
proportional to the temperature at which the created condensate disappears.
\begin{figure}[tbp]
\centering
\includegraphics[width=8cm]{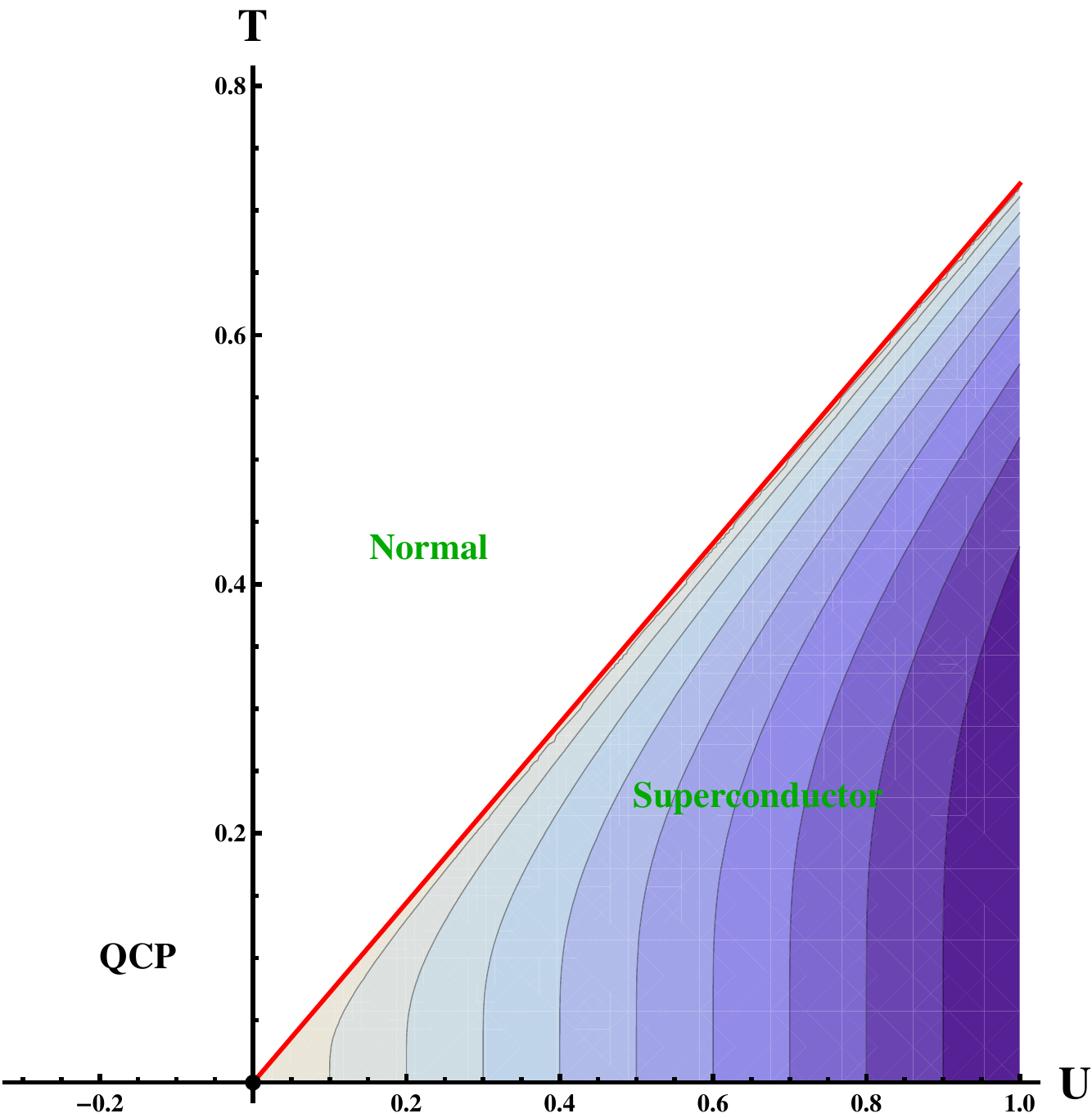}
\caption{Phase diagram of \ STI. Order parameter as function of chemical
potential and temperature near the quantum critical point at $T=0$, $\protect%
\mu =0$. The critical line is a strait line in mean field approximation.}
\end{figure}

The temperature dependence of the gap reads, see Fig. 4
\begin{equation}
\Delta \left( T\right) =2T\cosh ^{-1}\left( \frac{1}{2}\exp \frac{U}{2T}%
\right) \text{.}  \label{delta_mic}
\end{equation}
This it typical for chiral universality classes \cite{Sachdev,Rosenstein}.

It is interesting to compare this dependence with the conventional BCS for
transition at finite temperature, namely away from QCP. At zero temperature $%
\Delta \left( 0\right) /T_{c}=2\log 2$ $=\allowbreak 1.\,\allowbreak 39$
(within BCS - $1.76$), while near $T_{c}$ one gets $\Delta
/T_{c}=2^{3/2}\log ^{1/2}2\sqrt{1-t}=$ $2.\,\allowbreak 35\sqrt{1-t}$ (BCS -
$3.07\sqrt{1-t}$), where $t=T/T_{c}$. To describe the behavior of the STI in
inhomogeneous situations like the external magnetic field, boundaries,
impurities or junction with metals or other superconductors, it is necessary
to derive the effective theory in terms of the order parameter $\Delta
\left( \mathbf{r}\right) $, where $\mathbf{r}$ varies on the mesoscopic
scale.

\section{ Ginzburg - Landau effective theory and magnetic properties of the
superconductor near QCP}

\subsection{Coherence length and the condensation energy}

Using the well known Gor'kov method\cite{AGD}, the quadratic term of the
Ginzburg-Landau energy $F_{2}=\sum_{\mathbf{p}}\Delta _{p}^{\ast }\Gamma
\left( p\right) \Delta _{p}$ is obtained exactly from expanding the gap
equation to linear terms in $\Delta $ for arbitrary external momentum. The
result derived in Appendix B reads:
\begin{equation}
\Gamma \left( p\right) =-\frac{U}{4\pi \hbar ^{2}v_{F}^{2}}+\frac{\left\vert
p\right\vert }{16v_{F}\hbar ^{2}}\text{.}  \label{Gamma1}
\end{equation}%
The dependence on $\mathbf{p}$ is non-analytic and within our approximation
higher powers of $p$ do not appear. The second term is very different from
the quadratic term in the GL functional for conventional phase transitions
at finite temperature \cite{Landau} or even quantum phase transitions in
models without Weyl fermions \cite{Sachdev} and has a number of qualitative
consequences. Comparing the two terms in Eq.(\ref{Gamma1}), one obtains the
coherence length as a power of parameter $U\propto g-g_{c}$ describing the
deviation from criticality:
\begin{equation}
\xi \left( U\right) =\frac{\pi }{4}v_{F}\hbar U^{-\nu }\text{; \ \ \ \ }\nu
=1\text{.}  \label{coherence}
\end{equation}%
This is different from the dependence in non-chiral universality classes
that is\cite{Landau} $\xi \left( T\right) \infty \left( T_{c}-T\right)
^{-\nu },$ $\nu =1/2$ in mean field. Of course in the regime of critical
fluctuations this exponent is corrected in both non-chiral \cite{Landau} and
chiral\cite{Gat} universality classes.

Local terms in the GL energy density are also calculable exactly (within our
approximation, see Appendix C).%
\begin{equation}
f_{cond}=\frac{1}{4\pi \hbar ^{2}v_{F}^{2}}\left\{ -U\Delta ^{\ast }\Delta +%
\frac{2}{3}\left( \Delta ^{\ast }\Delta \right) ^{3/2}\right\} \text{.}
\label{Fcond}
\end{equation}%
It is quite nonstandard compared to customary quartic term $\left( \Delta
^{\ast }\Delta \right) ^{2}$ in conventional universality classes. The GL
equations in the homogeneous case for the condensate gives $\Delta
_{0}=U^{\beta }$ with critical exponent $\beta =1,$ different from the mean
field value $\beta =1/2$ for the $U\left( 1\right) $ universality class\cite%
{Landau}. The condensation energy density is $f_{0}=-\frac{1}{12\pi \hbar
^{2}v_{F}^{2}}U^{2-\alpha }$ with $\alpha =-1$. The free energy critical
exponent at QCP therefore is also different from the classical $\alpha =0$.

Having calculated both the local terms and the momentum dependence of the
quadratic term in the Ginzburg - Landau energy, one is ready to formulate
the GL energy in an inhomogeneous situations including magnetic field.

\subsection{GL equations in the presence of magnetic field}

In view of the local gauge invariance principle, replacing the momentum by a
covariant derivative, the gradient term of the GL energy becomes
\begin{equation}
F_{grad}=\int d^{2}\mathbf{r}\frac{1}{16v_{F}\hbar }\Delta ^{\ast }\left(
\mathbf{r}\right) \sqrt{\left( -i\partial _{i}-\frac{e^{\ast }}{c\hbar }%
A_{i}\left( \mathbf{r}\right) \right) ^{2}}\Delta \left( \mathbf{r}\right)
\text{.}  \label{Fgrad}
\end{equation}%
This should be supplemented by the condensation energy Eq.(\ref{Fcond}) and
magnetic energy Eq.(\ref{Hmag}). The GL equations are obtained by
minimization with respect to 2D order parameter and 3D vector potential. In
the present case the equation for the order parameter is nonlocal and
nonanalytic:
\begin{equation}
\left\{ \xi \sqrt{\left( -i\partial _{i}-\frac{e^{\ast }}{\hbar c}%
A_{i}\right) ^{2}}-1\right\} \Delta +\frac{\Delta }{U}\left( \Delta ^{\ast
}\Delta \right) ^{1/2}=0\text{.}  \label{GLeq}
\end{equation}%
The supercurrent in the Maxwell equation,%
\begin{equation}
\frac{c}{4\pi }\mathbf{\nabla }\times \mathbf{B}=\mathbf{J}\left( \mathbf{r}%
\right) \delta \left( z\right) ,  \label{Maxwell}
\end{equation}%
is also nonlocal: $J_{i}\left( \mathbf{r}\right) =\frac{1}{c}\frac{\delta F}{%
\delta A_{i}\left( \mathbf{r}\right) }$.

\subsection{Upper critical field and the magnetization curve}

The upper critical field is found from the spectrum of the gradient term
operator in Eq.(\ref{GLeq}). The lowest eigenvalue of the operator for
homogeneous induction $\mathbf{B}=\left\{ 0,0,B\right\} $ is $\xi \sqrt{%
e^{\ast }B/c\hbar }$ (the eigenvalue of the square root of an operator is a
square root of the eigenvalue) and therefore the bifurcation occurs at%
\begin{equation}
H_{c2}=\frac{\Phi _{0}}{2\pi \xi ^{2}}\text{,}  \label{Hc2}
\end{equation}%
with the coherence length $\xi $ found in Section II, Eq.(\ref{coherence}).
The formula is the same as in a more customary situation despite the fact
that the coherence length has a different origin and different critical
exponent at QCP.

Near $H_{c2}$ the Abrikosov hexagonal lattice is formed. Its energy density
is approximated well using the lowest Landau level (LLL) approximation: $%
\Delta \left( \mathbf{r}\right) =\Delta _{A}\varphi \left( \mathbf{r}\right)
$, where the Abrikosov hexagonal lattice function $\varphi $ is normalized
by $\left\langle \left\vert \varphi \left( \mathbf{r}\right) \right\vert
^{2}\right\rangle =1$ ($\left\langle ...\right\rangle $ denotes here the
space average). The strength of the condensate is determined by minimizing
the energy (magnetic energy can be neglected):
\begin{eqnarray}
\left\langle f\right\rangle &=&\frac{\left\vert \Delta _{A}\right\vert ^{2}}{%
16v_{F}\hbar }\left\langle \varphi ^{\ast }\left( \sqrt{\left( -i\partial
_{i}-\frac{e^{\ast }}{c\hbar }A_{i}\left( \mathbf{r}\right) \right) ^{2}}-%
\frac{4U}{\pi v_{F}\hbar }\right) \varphi \right\rangle  \notag \\
&&+\frac{\left\vert \Delta _{A}\right\vert ^{3}}{6\pi \hbar ^{2}v_{F}^{2}}%
\left\langle \left\vert \varphi \right\vert ^{3}\right\rangle  \label{f} \\
&=&-\frac{\left\vert \Delta _{A}\right\vert ^{2}}{32v_{F}\hbar }\sqrt{\frac{%
e^{\ast }}{c\hbar }}H_{c2}^{1/2}\left( 1-H/H_{c2}\right) +\frac{\beta
_{3}\left\vert \Delta _{A}\right\vert ^{3}}{6\pi \hbar ^{2}v_{F}^{2}}\text{.}
\notag
\end{eqnarray}%
The number $\beta _{3}=\left\langle \left\vert \varphi \right\vert
^{3}\right\rangle =1.07$ is analogous to $\beta _{A}$ for usual fourth power
GL energy. The optimal $\Delta _{A}$ at external field $H$ close to $H_{c2}$
is:
\begin{equation}
\left\vert \Delta _{A}\right\vert =\frac{U}{2\sqrt{2\pi }\beta _{3}}\left(
1-H/H_{c2}\right) ^{\sigma }\text{; \ \ \ }\sigma =1\text{.}  \label{psiLLL}
\end{equation}%
This exponent for the transition on the $H_{c2}$ line is different from the
ordinary Abrikosov lattice\cite{Abrikosov} for which $\sigma =1/2$.

The magnetization is calculated from the averaged energy density for the
optimal $\Delta _{A}$ given in Eq.(\ref{psiLLL}) is ($B\simeq H$):
\begin{equation}
f\left( B\right) =-\frac{\sqrt{2\pi }}{3\cdot 2^{10}\beta _{3}}\frac{UH_{c2}%
}{\Phi _{0}}\left( 1-H/H_{c2}\right) ^{3}\text{.}  \label{f_B}
\end{equation}%
The dependence is quadratic,
\begin{equation}
M=-\frac{\pi ^{3/2}}{2^{7}\sqrt{2}\beta _{3}}\frac{U}{\Phi _{0}}\left(
1-H/H_{c2}\right) ^{\tau }\text{, }\tau =2\text{, }  \label{magn}
\end{equation}%
that should be contrasted with the usual linear dependence\cite{Abrikosov}, $%
\tau =1$. For smaller fields the vortex lattice becomes less dense and
eventually the LLL approximation \cite{Li} breaks down. However, since the
superconductivity is confined to an atomic-width layer, there is no $H_{c1},$
and at small fields vortices become independent. Consequently the parabolic
increase is halted and perfect diamagnetism appears only at $H=0$. Under
these conditions we turn to a single vortex solution next.
\begin{figure}[tbp]
\centering
\includegraphics[width=8cm]{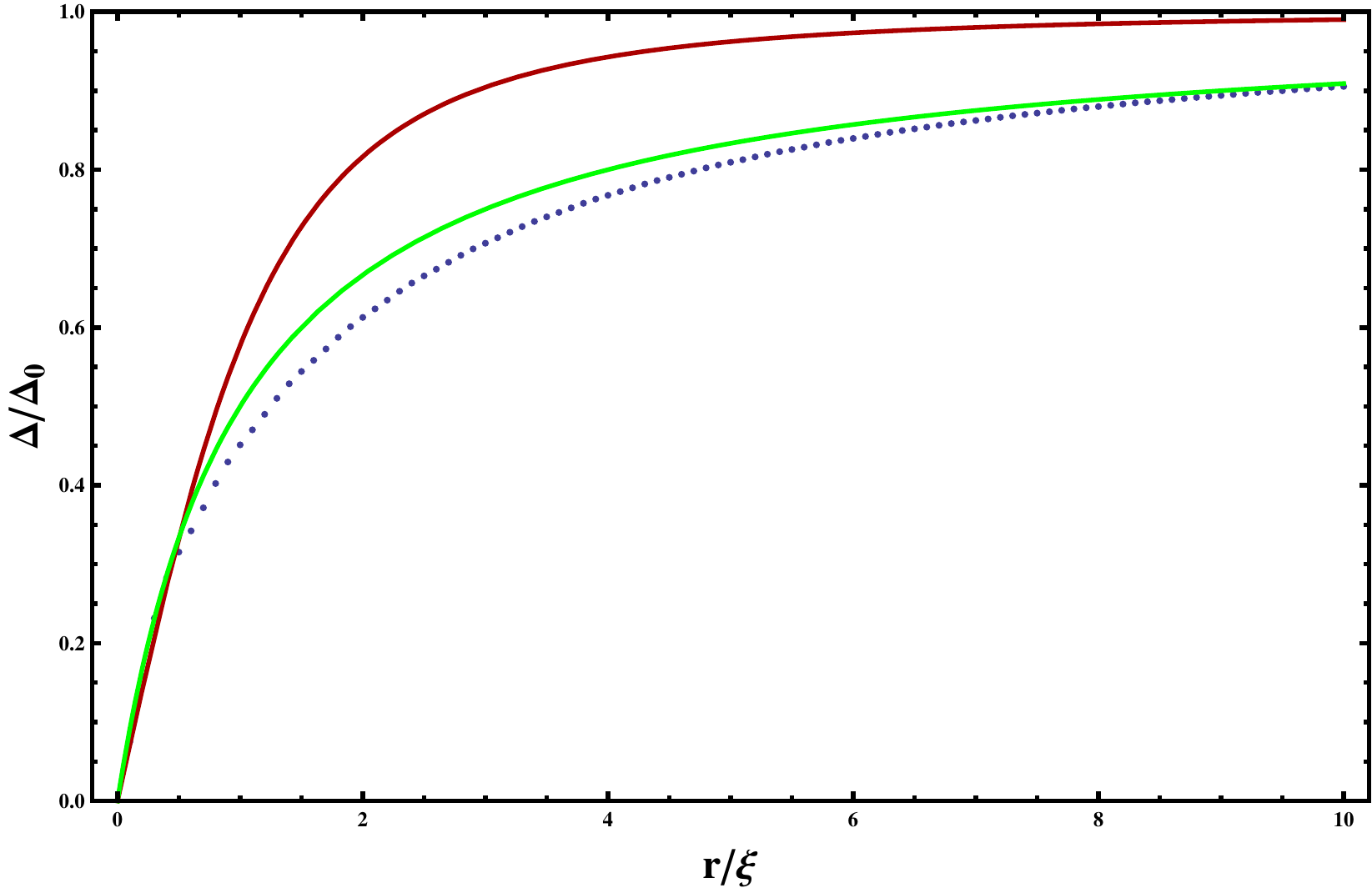}
\caption{Vortex core structure near QCP. Order parameter in units of the
bulk gap $\Delta _{0}$ as function of distance from the center in units of
coherence length $\protect\xi $. The blue line is the approximate formula,
while the red line is the usual Abrikosov vortex profile.}
\end{figure}

\subsection{Core structure of a single vortex}

The single vortex solution for the order parameter can be found using the
rotational symmetry in polar coordinates: $\Delta =Uf\left( r\right)
e^{i\phi }$ with the homogeneous condensate value $\Delta =U$ found in
Section II, so that at large distances the dimensionless order parameter $%
f\left( r\right) \rightarrow 1$. At the center of the vortex $f$ vanishes.
The effects of the magnetic field, other than the phase rotation, are small
in this extreme type II case of a surface superconductor \cite{Abrikosov}.
In this case the GL equation Eq.(\ref{GLeq}), using the coherence length $%
\xi $, Eq.(\ref{coherence}) as unit of length, $\mathbf{r}\rightarrow \xi
\mathbf{r\,}$, takes the form:%
\begin{equation}
\left( \sqrt{\widehat{L}}-1\right) f\left( r\right) +f\left( r\right) ^{2}=0%
\text{.}  \label{dimensionlesseq}
\end{equation}%
The operator $\widehat{L}\equiv -\partial _{r}^{2}-\frac{1}{r}\partial _{r}+%
\frac{1}{r^{2}}$ has Bessel functions as its eigenvectors,
\begin{equation}
\left( -\partial _{r}^{2}-\frac{1}{r}\partial _{r}+\frac{1}{r^{2}}\right)
J_{1}\left( \alpha r\right) =\alpha ^{2}J_{1}\left( \alpha r\right) \text{.}
\label{Bessel}
\end{equation}

\begin{figure*}[tbp]
\begin{center}
\includegraphics[width=16cm]{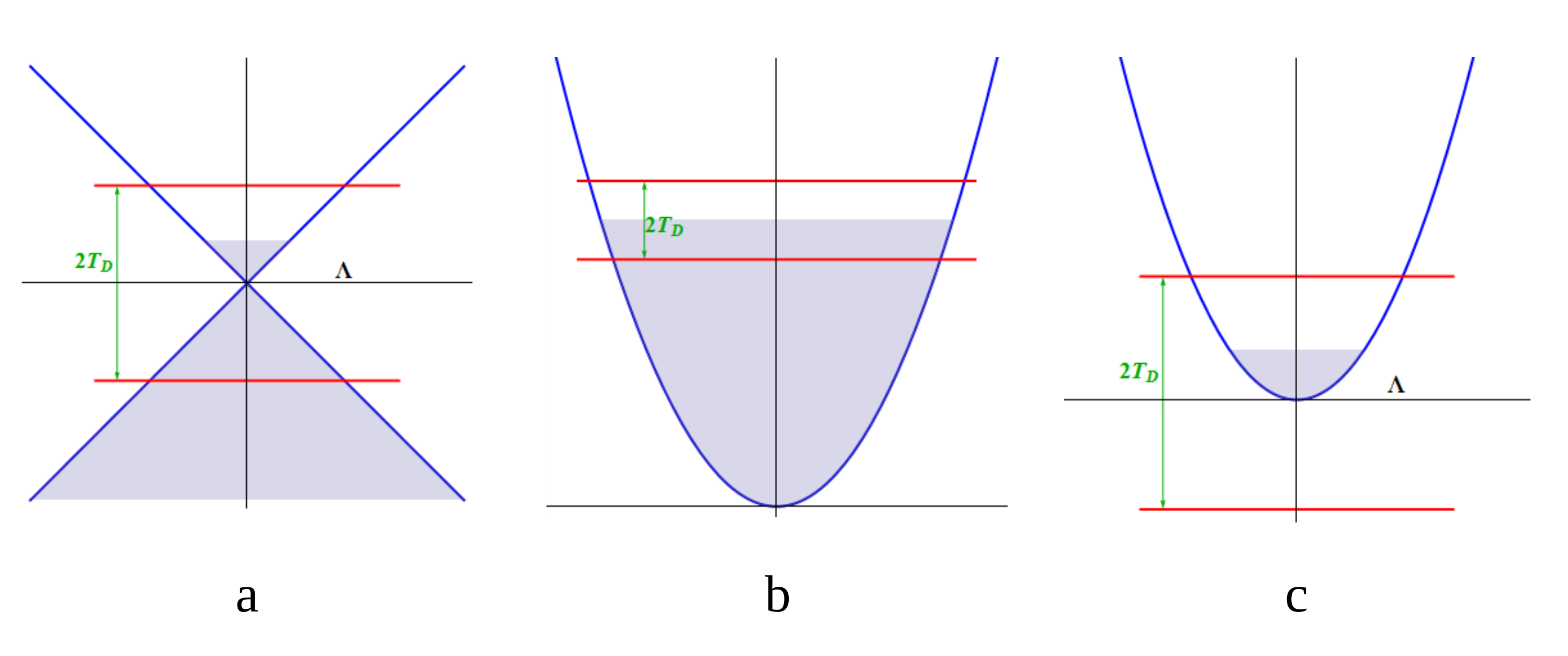}
\end{center}
\par
\vspace{-0.5cm}
\caption{Schematic picture of the band reconstruction due to phonon pairing
in three different 2D fermionic systems. a. Weyl semi-metal. b. BEC with
parabolic dispersion law. c. Classic BCS.}
\end{figure*}
Looking for the solution expanded in full set of these functions for all $%
\alpha $ satisfying our boundary conditions (Hankel transform) in the form%
\begin{equation}
f\left( r\right) =1-\int_{\alpha =0}^{\infty }\alpha F\left( \alpha \right)
J_{1}\left( \alpha r\right) \text{,}  \label{CoefF}
\end{equation}%
the equation becomes
\begin{equation}
\int_{\alpha =0}^{\infty }F\left( \alpha \right) \alpha \left( \alpha
+1\right) J_{1}\left( \alpha r\right) =\left( \int_{\beta =0}^{\infty }\beta
F\left( \beta \right) J_{1}\left( \beta r\right) \right) ^{2}\text{.}
\label{Feq}
\end{equation}%
To obtain an iterative form we multiply by $rJ_{1}\left( \gamma r\right) ,$
and integrating over $r$ using explicit formulas \cite{Auluck} given in
Appendix D results in
\begin{equation}
F\left( \gamma \right) =\frac{1}{\pi \left( \gamma +1\right) }\int_{\alpha
,\beta =0}^{\infty } \frac{F\left( \alpha \right) F\left( \beta \right)
(\alpha ^{2}+\beta ^{2}-\gamma ^{2})}{\sqrt{\left( \gamma ^{2}-\left( \alpha
-\beta \right) ^{2}\right) \left( \left( \alpha +\beta \right) ^{2}-\gamma
^{2}\right) }}\text{.}  \label{iteration}
\end{equation}

The iteration converges very fast with the result presented in Fig. 5
(dots). The asymptotic at small $r$ is linear, $f\left( r\right) =r,$ while
at large $r$ as expected it approaches the "bulk" value $f\left( r\right)
\rightarrow 1,$ and can be approximated by a formula $f\left( r\right) =%
\frac{r}{r+1}$ (green curve in Fig.5), simpler then the usual interpolation
formula, $f\left( r\right) =\frac{r}{\sqrt{1+2r^{2}}}$ (orange curve in
Fig.5). One observes that the relaxation of the order parameter away of the
center of the vortex is much slower in STI..

\section{Discussion and conclusions}

\subsection{Comparison of renormalization of the coupling with BEC and BCS
in 2D.}

Let us contrast the coupling renormalization in a 2D Weyl semi-metal with
momentum cutoff $\Lambda $ (for definiteness one can assume the phonon
mechanism so that $\Lambda $ is the Debye cutoff $T_{D}$, under condition
that the deviation from the Weyl point $\mu <<T_{D}$) with that in a 2D
parabolic band, $E_{p}=\frac{p^{2}}{2m^{\ast }}$.The renormalized coupling,
Eqs.(\ref{g_c}\ref{gren}), can be written in the form%
\begin{equation}
\frac{1}{g_{ren}}=\frac{1}{g}-\frac{\Lambda }{4\pi \hbar ^{2}v_{F}}\text{,}
\label{renormalization}
\end{equation}%
where $g_{ren}\equiv -4\pi \hbar ^{2}v_{F}^{2}/U$. The linear
renormalization (rather than the customary logarithmic cutoff dependence) of
$\frac{1}{g}$ in Weyl semi-metal is pertinent to the so-called "chiral
universality classes" that sometimes appear in description of quantum
critical points in 2D\cite{Sachdev}. It corresponds to finite coupling $%
g_{c} $ fixed points in Eqs.(\ref{gren},\ref{g_c}). This is the main
difference from the more conventional cases that are briefly summarized next.

Within the parabolic case two cases are generally distinguished\cite{Stoof}:
the BCS, where the chemical potential $\mu $ is well above the bottom of the
band, see Fig. 6, so that $T_{D}<<\mu $ (like in metallic superconductors),
and the BEC when $\Lambda ^{2}/2m^{\ast }>>\mu $, (like in cold atoms).

In BEC, that is closer to the STI considered here, the gap equation reads:%
\begin{eqnarray}
\frac{1}{g^{BEC}}&=&\frac{1}{2\pi }\int_{k=\hbar /L}^{\Lambda }\frac{k}{%
\sqrt{\left( k^{2}/2m^{\ast }-\mu \right) ^{2}+\Delta ^{2}}}  \notag \\
&=&\frac{m^{\ast }}{4\pi \hbar ^{2}}\log \frac{\Lambda ^{2}}{m^{\ast }\left(
\sqrt{\mu _{ren}^{2}+\Delta ^{2}}-\mu _{ren}\right) }\text{.}  \label{BECgap}
\end{eqnarray}%
Here $L$ is an infrared cutoff (needed in 2D) that is incorporated in $\mu
_{ren}=\mu -\frac{\hbar ^{2}}{2L^{2}m^{\ast }}$. The corresponding
renormalized coupling depends on the reference (normalization) point $%
E_{ren} $ \cite{Shankar}:%
\begin{equation}
\frac{1}{g_{ren}^{BEC}}=\frac{1}{g^{BEC}}-\frac{m^{\ast }}{4\pi \hbar ^{2}}%
\log \frac{\Lambda ^{2}}{m^{\ast }E_{ren}}\text{.}  \label{renBEC}
\end{equation}%
In terms of this coupling the theory becomes cutoff independent. For
example, the gap equation reads:%
\begin{equation}
\frac{1}{g_{ren}^{BEC}}=\frac{m^{\ast }}{4\pi \hbar ^{2}}\log \frac{E_{ren}}{%
\sqrt{\mu _{ren}^{2}+\Delta ^{2}}-\mu _{ren}}\text{.}  \label{ren_gap_BEC}
\end{equation}

\begin{table*}[tph]
\caption{Critical exponents of the chiral universality class of the TI QCP}
\begin{center}
\renewcommand\arraystretch{1.5}
\begin{tabularx}{\textwidth}{p{4cm}XXXX}
\hline
\hline
\text{\textbf{critical exponent}} & \text{order parameter} & \text{coherence
length} & \text{energy} & \text{temperature} \\
\hline
\text{QCP }$U_{1}\left( 1\right)$ \text{definition} & $\Delta \propto
U^{\beta }$ & $\xi =U^{-\nu }$ & $f\propto U^{2-\alpha }$ & $T_{c}\propto U^{z\nu }$
\\
\text{meanfield value} & $\beta =1$ & $\nu =1$ & $\alpha =-1$ & $z\nu =1$ \\
\text{classical }$U\left( 1\right)$ \text{definition} & $\Delta \propto
\left( T_{c}-T\right) ^{\beta }$ & $\xi \propto \left( T_{c}-T\right) ^{-\nu }$
& $f\propto \left( T_{c}-T\right) ^{2-\alpha }$ & -- \\
\text{mean field value} & $\beta =\frac{1}{2}$ & $\nu =\frac{1}{2}$ & $\alpha =0$
& --\\
\hline
\hline
\end{tabularx}
\end{center}
\end{table*}

\begin{table*}[tph]
\caption{Critical exponents of the chiral universality class of the
Abrikosov transition in external magnetic field at QCP.}
\begin{center}
\renewcommand\arraystretch{1.5}
\begin{tabularx}{\textwidth}{XXX}
\hline
\hline
\textbf{critical exponent} & magnetization & OP magnetic\\
\hline
QCP $U_{1}\left( 1\right)$  definition & $M\propto \left(
H_{c2}-H\right) ^{\tau }$ & $\Delta _{A}\propto \left( H_{c2}-H\right)
^{\sigma }$ \\
mean field value & $\tau =2$ & $\sigma =1$ \\
Abrikosov lattice definition & $M\propto \left( H_{c2}-H\right) ^{\tau}$ & $\Delta _{A}\propto \left( H_{c2}-H\right) ^{\sigma }$ \\
mean field value & $\tau =1$ & $\sigma =\frac{1}{2}$\\
\hline
\hline
\end{tabularx}
\end{center}
\end{table*}

In BCS the gap equation, under the simplifying conditions $\mu
>>T_{D}>\Delta $ (the dispersion relation near the Fermi level can be
approximated by a "flat" one\cite{AGD}), is
\begin{eqnarray}
\frac{1}{g^{BCS}}&=&\frac{1}{2\pi }\int_{k=\sqrt{2m^{\ast }\left( \mu
-T_{D}\right) }}^{\sqrt{2m^{\ast }\left( \mu +T_{D}\right) }}\frac{k}{\sqrt{%
\left( k^{2}/2m^{\ast }-\mu \right) ^{2}+\Delta ^{2}}}  \notag \\
&\simeq& \frac{m^{\ast }}{2\pi \hbar ^{2}}\log \frac{2T_{D}}{\Delta }\text{.}
\label{bareBEC}
\end{eqnarray}%
The renormalized coupling is again dependent on an arbitrarily chosen
normalization scale, $E_{ren}$,%
\begin{equation*}
\frac{1}{g_{ren}^{BCS}}=\frac{1}{g^{BCS}}-\frac{m^{\ast }}{2\pi \hbar ^{2}}%
\log \frac{T_{D}}{E_{ren}}\text{.}
\end{equation*}

In both parabolic cases the coupling is logarithmically "running" towards
weak coupling\cite{Shankar} $g^{BCS}\rightarrow g_{c}=0$ (marginally
irrelevant or asymptotically free) at large $\Lambda $. In the Weyl semimetal
(where the dispersion is linear) with local interaction the criticality
appears at small $U$ when $g$ approaches finite value $g_{c}$. Despite the
fact that the UV cutoff does not appear logarithmically, the theory is still
renormalizable\cite{Rosenstein} and any physical quantity can be expressed
via renormalized coupling $U$.

\subsection{Experimental feasibility of the surface superconductivity due to
phonon exchange}

To estimate the pairing efficiency due to phonons, one should rely on recent
studies of surface phonons in TI \cite{DasSarma}. The coupling constant in
the Hamiltonian, Eq.(\ref{Hamiltonian}), is obtained from the exchange of
acoustic (Rayleigh) surface phonons $g=\lambda v_{F}^{2}\hbar ^{2}/2\pi \mu $%
, where $\lambda $ is the dimensionless effective electron - electron
interaction constant of order $0.1$ (somewhat lower values are obtained in
ref.\cite{Guinea}). It was shown in ref. \cite{DasSarma} that at zero
temperature the ratio of $\lambda $ and $\mu $ is constant with well defined
$\mu \rightarrow 0$ limit with value $g=0.23$ $eV\ nm^{2}$ for $v_{F}\approx
7\ \cdot 10^{5}m/s$ (for $Bi_{2}Se_{3}$). The critical coupling constant $%
g_{c}$, Eq.(\ref{g_c}), can be estimated from the Debye cutoff $T_{D}=200K$
determining the momentum cutoff $\Lambda =T_{D}/c_{s}$, where $c_{s}$ is the
sound velocity. Taking value to be $c_{s}=2\cdot 10^{3}m/s$ (for $%
Bi_{2}Se_{3}$), one obtains $g_{c}=4\pi v_{F}c_{s}\hbar ^{2}/\,T_{D}=0.20$ $%
eV$ $nm^{2}$. Therefore the stronger superconductivity, $g>g_{c}$, is
realized (see Fig.3 and case (i) of Section IIC, $U>0$). Note that the
superconductivity appears even for $0<g<g_{c}$ ($U>0$ in Fig. 3), although,
as discussed in Section IIC case (ii), it is weaker.

Of course the Coulomb repulsion might weaken or even overpower the effect of
the attraction due to phonons, so that superconductivity does not occur. In
TI like $Bi_{2}Se_{3}$ however, the dielectric constant is very large $%
\varepsilon =50$, so that the Coulomb repulsion is weak. Moreover it was
found in graphene (that has identical Coulomb interaction), that although
the semi-metal does not screen \cite{CastroNeto}, the effects of the Coulomb
coupling are surprisingly small, even in leading order in perturbation
theory.

Superconductivity was observed in otherwise nonsuperconducting TIs $%
Bi_{2}Te_{3}$ and $Bi_{2}Se_{3}$. It was noticed very recently\cite%
{Luo,Koren} that $Bi$ nanoclusters naturally aggregate on the surface of $%
Bi_{2}Te_{3}$ thin film and an explanation was put forward that the
nanoclusters become superconducting and induce surface superconductivity in
TI by the proximity effect. We speculate that the nanoclusters are not
superconducting and their role might be to screen the Coulomb repulsion.

In this paper we focused on the qualitatively distinct case of Weyl fermions
with small chemical potential. Although in the original proposal of TI in
materials \cite{Zhang1} the chemical potential was zero, in experiments one
finds often that the Dirac point is shifted away from the Fermi surface by a
significant fraction of $eV$ \cite{Zhang}. There are however experimental
methods to shift the location of the point by doping, gating, pressure etc.%
\cite{bias}. Note that a reasonable electron density of $n=3\cdot
10^{11}cm^{-2}$ in $Bi_{2}Te_{3}$ already conforms to the requirement that
chemical potential $\mu =\sqrt{n}\hbar v_{F}/2\pi =100K$ \ is smaller that
the Debye cutoff energy $T_{D}=200K$.

\subsection{Conclusions}

We have studied the s-wave pairing on the surface of 3D topological
insulator. The noninteracting system is characterized by (nearly) zero
density of states on the 2D Fermi manifold. It degenerates into a point when
the chemical potential coincides with the Weyl point of the surface states
as in the original proposal for a major class of such materials\cite{Zhang1}%
. The pairing attraction (the most plausible candidate being surface
phonons) therefore has two tasks in order to create the superconducting
condensate. The first is to create a pair of electrons (that in the present
circumstances means creating two holes as well) and the second is to pair
them. To create the charges does not cost much energy since the spectrum of
the Weyl semimetal is gapless (massless relativistic fermions); this is
effective as long as the coupling $g$ is larger than the critical $g_{c}$,
see Eq.(\ref{g_c}). The situation is more reminiscent of the creation of the
chiral condensate in relativistic massless four - fermion theory (a 2D
version\cite{Rosenstein} was recently contemplated for graphene \cite%
{CastroNeto,Katsnelson}) than to the BCS or even BEC in condensed matter
systems with parabolic dispersion law. Due to the special
"ultra-relativistic" nature of the pairing transition at zero temperature as
a function of parameters like the pairing interaction strength is unusual:
even the mean field critical exponents are different from the standard ones
that generally belong to the $U\left( 1\right) $ class of second order phase
transitions.

To summarize, we calculated, using the Gor'kov theory, the phase diagram of
the superconducting transition at arbitrary chemical potential $\mu $,
effective coupling energy $U$ and temperature $T$, see Figs.2,3. The quantum
($T=0$) critical point appears at $\mu =0$, $U=0$ and belongs the $%
U_{1}\left( 1\right) $ \textit{chiral} universality class (the subscript
denotes number of massless fermions at QCP) according to classification in
\cite{Gat,Sachdev}. The critical exponents are summarized in Table 1 for the
"static" exponents and Table 2 for response to temperature and magnetic
field (gauge coupling). The Ginzburg - Landau effective theory near the QCP,
Eq.(\ref{GLeq}), was derived and is rather unusual. The magnetization curve
near $H_{c2}$ due to vortex lattice is parabolic rather than linear. This
might be important for experimental identification of QCP. The vortex core
structure was determined, see Fig. 5, has some peculiarities that can be
tested directly.

\textit{Acknowledgements.} We are indebted to C. W. Luo, J. J. Lin and W.B.
Jian for explaining details of experiments, and T. Maniv and M. Lewkowicz
for valuable discussions. Work of B.R. and D.L. was supported by NSC of
R.O.C. Grants No. 98-2112-M-009-014-MY3 and MOE ATU program. The work of
D.L. also is supported by National Natural Science Foundation of China (No.
11274018),

\section{Appendix A. Integrals and sums for gap equation}

The bubble integral in the gap equation Eq.(\ref{gap eq}), at finite
temperature can be written as:
\begin{eqnarray}
b &=&\frac{T}{2}\sum\limits_{n,\mathbf{p}}\left\{ \frac{1}{\Delta ^{2}+\hbar
^{2}\omega _{n}^{2}+\left( v_{F}p+\mu \right) ^{2}}\right.  \label{Bub1} \\
&&\left. +\frac{1}{\Delta ^{2}+\hbar ^{2}\omega _{n}^{2}+\left( v_{F}p-\mu
\right) ^{2}}\right\} \text{.}  \notag
\end{eqnarray}%
At zero temperature after integration over frequencies, it becomes
(summation over momenta is replaced by integral with momentum cutoff $%
\Lambda $ in polar coordinates)
\begin{eqnarray}
b &=&\frac{1}{8\pi \hbar ^{2}}\int_{p=0}^{\Lambda }p\left\{ \frac{1}{\sqrt{%
\Delta ^{2}+\left( v_{F}p+\mu \right) ^{2}}}\right.  \label{Bub2} \\
&&\left. +\frac{1}{\sqrt{\Delta ^{2}+\left( v_{F}p-\mu \right) ^{2}}}%
\right\} \text{.}
\end{eqnarray}%
The integral is readily performed and expanded in $1/\Lambda $
\begin{eqnarray}
b &=&\frac{1}{8\pi \hbar ^{2}v_{F}^{2}}\left\{ \sqrt{\Delta ^{2}+\left(
v_{F}\Lambda +\mu \right) ^{2}}+\sqrt{\Delta ^{2}+\left( v_{F}\Lambda -\mu
\right) ^{2}}+\right.  \label{Bub3} \\
&&\mu \log \frac{\left( \sqrt{\Delta ^{2}+\mu ^{2}}+\mu \right) \left(
v_{F}\Lambda -\mu +\sqrt{\Delta ^{2}+\left( v_{F}\Lambda -\mu \right) ^{2}}%
\right) }{\left( \sqrt{\Delta ^{2}+\mu ^{2}}-\mu \right) \left( v_{F}\Lambda
+\mu +\sqrt{\Delta ^{2}+\left( v_{F}\Lambda +\mu \right) ^{2}}\right) }
\notag \\
&&\left. -2\sqrt{\Delta ^{2}+\mu ^{2}}\right\}  \notag \\
&\simeq &\frac{1}{4\pi \hbar ^{2}v_{F}^{2}}\left\{ v_{F}\Lambda \ -\sqrt{%
\Delta ^{2}+\mu ^{2}}+\frac{\mu }{2}\log \frac{\mu +\sqrt{\Delta ^{2}+\mu
^{2}}}{\sqrt{\Delta ^{2}+\mu ^{2}}-\mu }\right\}  \notag \\
&&+O\left( \frac{1}{\Lambda }\right) \text{.}  \notag
\end{eqnarray}

\ At finite temperature, using the sum,
\begin{equation}
T\sum\limits_{n}\left( \omega _{n}^{2}+m^{2}\right) ^{-1}=\frac{\tanh \left[
m/\left( 2T\right) \right] }{2m}\text{,}  \label{sum_temp}
\end{equation}%
one obtains%
\begin{eqnarray}
B &=&\frac{1}{8\pi }\int_{p=0}^{\Lambda }p\left\{ \frac{\tanh \left[ \sqrt{%
\Delta ^{2}+\left( v_{F}p+\mu \right) ^{2}}/\left( 2T\right) \right] }{\sqrt{%
\Delta ^{2}+\left( v_{F}p+\mu \right) ^{2}}}\right.  \label{B_temp} \\
&&\left. +\frac{\tanh \left[ \sqrt{\Delta ^{2}+\left( v_{F}p-\mu \right) ^{2}%
}/\left( 2T\right) \right] }{\sqrt{\Delta ^{2}+\left( v_{F}p-\mu \right) ^{2}%
}}\right\} \text{.}
\end{eqnarray}%
For $\mu =0$ it simplifies,
\begin{eqnarray}
b &=&\frac{1}{4\pi \hbar ^{2}}\int_{p=0}^{\Lambda }p\frac{\tanh \left[ \sqrt{%
\Delta ^{2}+v_{F}^{2}p^{2}}/\left( 2T\right) \right] }{\sqrt{\Delta
^{2}+v_{F}^{2}p^{2}}}  \label{B_zero_temp} \\
&=&\frac{1}{4\pi \hbar ^{2}v_{F}^{2}}\left\{ v_{F}\Lambda -2T\log \left[
2\cosh \left( \frac{\Delta }{2T}\right) \right] \right\} \text{,}  \notag
\end{eqnarray}%
This was used in Eq.(\ref{gapren_T}). For $\Delta =0$ and $\mu \not=0$%
\begin{eqnarray}
b &=&\frac{1}{8\pi \hbar ^{2}}\int_{p=0}^{\Lambda }p\left\{ \frac{\tanh %
\left[ \left\vert v_{F}p+\mu \right\vert /\left( 2T_{c}\right) \right] }{%
\left\vert v_{F}p+\mu \right\vert }\right.  \label{crit_temp} \\
&&\left. +\frac{\tanh \left[ \left\vert v_{F}p-\mu \right\vert /\left(
2T_{c}\right) \right] }{\left\vert v_{F}p-\mu \right\vert }\right\} \text{.}
\end{eqnarray}

\section{Appendix B. Critical line and condensate}

The dependence of the gap on chemical potential given in Eq.(\ref{gap}). For
positive $U$ and $\mu <<\Delta $ the formula can be expanded as
\begin{equation}
U=\sqrt{\Delta ^{2}+\mu ^{2}}-\frac{\mu }{2}\log \frac{\sqrt{\Delta ^{2}+\mu
^{2}}+\mu }{\sqrt{\Delta ^{2}+\mu ^{2}}-\mu }=\Delta -\frac{\mu ^{2}}{%
2\Delta }  \label{B1}
\end{equation}%
from which Eq.(\ref{delta_pos}) follows. In the case of $U=0$ the equation
becomes homogeneous:%
\begin{equation}
\sqrt{\Delta ^{2}+\mu ^{2}}=\frac{\mu }{2}\log \frac{\sqrt{\Delta ^{2}+\mu
^{2}}+\mu }{\sqrt{\Delta ^{2}+\mu ^{2}}-\mu }\rightarrow \frac{\Delta }{\mu }%
=0.663\text{.}  \label{B2}
\end{equation}%
In the negative $U$ case $\Delta $ is exponentially small (so that $\mu
>>\Delta $) and
\begin{eqnarray}
U &=&\sqrt{\Delta ^{2}+\mu ^{2}}-\frac{\mu }{2}\log \frac{\sqrt{\Delta
^{2}+\mu ^{2}}+\mu }{\sqrt{\Delta ^{2}+\mu ^{2}}-\mu }  \label{B3} \\
&\simeq &\mu -\mu \log \frac{\mu }{\Delta }\text{,}  \notag
\end{eqnarray}%
from which Eq.(\ref{delta_neg}) follows.

For critical temperature for arbitrary $\mu $ is obtained as the $\Delta
\rightarrow 0$ limit of the gap equation Eq.(\ref{gapren_T}).
\begin{eqnarray}
U &=&\frac{v_{F}}{2}\int_{p=0}^{\Lambda }p\left\{ \frac{\tanh \left[
\left\vert v_{F}p+\mu \right\vert /\left( 2T_{c}\right) \right] }{\left\vert
v_{F}p+\mu \right\vert }\right.  \label{B4} \\
&&\left. +\frac{\tanh \left[ \left\vert v_{F}p-\mu \right\vert /\left(
2T_{c}\right) \right] }{\left\vert v_{F}p-\mu \right\vert }-\frac{2}{%
\left\vert p\right\vert }\right\} \text{.}
\end{eqnarray}%
This is presented in Fig. 3.

\section{Appendix C. Derivation of the GL energy for at QCP}

\subsection{Linear term in GL equation for arbitrary momentum $p$.}

Expanding the right hand side of Eq.(\ref{gap eq}) to linear term, the
expression for the kernel can be written as a trace:
\begin{eqnarray}
\Gamma &=&\frac{1}{2}tr\left\{ \sum\limits_{\omega q}\sigma ^{y}D_{\omega
q}^{t}\sigma ^{y}D_{\omega ,q-p}+\frac{1}{g}I\right\}  \label{C1} \\
&=&\frac{1}{g}-\sum\limits_{\omega q}\frac{\hbar ^{2}\omega
^{2}-v_{F}^{2}p\cdot q+v_{F}^{2}q^{2}}{\left( \hbar ^{2}\omega
^{2}+v_{F}^{2}q^{2}\right) \left( \hbar ^{2}\omega ^{2}+v_{F}^{2}\left\vert
\mathbf{q-p}\right\vert ^{2}\right) }.  \notag
\end{eqnarray}%
Integrating over $\omega $ (at zero temperature) one obtains%
\begin{eqnarray}
\Gamma &=&-\frac{U}{4\pi \hbar ^{2}v_{F}^{2}}-\frac{1}{2v_{F}}\sum\limits_{q}%
\frac{p\cdot q-p^{2}}{q\left( q-p\right) ^{2}+q^{2}\left\vert \mathbf{q-p}%
\right\vert }  \label{C2} \\
&=&-\frac{1}{8\pi ^{2}\hbar ^{2}v_{F}}\int_{q,\phi }\frac{pq\cos \phi -p^{2}%
}{\left\vert \mathbf{q-p}\right\vert ^{2}+q\left\vert \mathbf{q-p}%
\right\vert }-\frac{U}{4\pi \hbar ^{2}v_{F}^{2}}  \notag
\end{eqnarray}%
where%
\begin{equation*}
\left\vert \mathbf{q-p}\right\vert ^{2}=q^{2}+p^{2}-2pq\cos \phi \text{.}
\end{equation*}%
The integral is homogeneous in momentum and therefore is linear in $%
p=\left\vert \mathbf{p}\right\vert $ and one arrives at Eq.(\ref{Gamma1}).

\subsection{\ Local terms in GL equation and energy}

For $p=0$ the gap equation Eq.(\ref{gapeq1}) reads%
\begin{equation}
\frac{\Delta }{4\pi \hbar ^{2}v_{F}^{2}}\left( -U+\sqrt{\Delta ^{\ast
}\Delta }\right) =0\text{,}  \label{C3}
\end{equation}%
This is obtained from the energy functional%
\begin{equation}
F=\frac{1}{4\pi \hbar ^{2}v_{F}^{2}}\int d^{2}\mathbf{r}\left\{ -U\Delta
^{\ast }\Delta +\frac{2}{3}\left( \Delta ^{\ast }\Delta \right)
^{3/2}\right\} \text{.}  \label{C4}
\end{equation}

\section{Appendix D. Single vortex.}

The basic integral of the Hankel transform is
\begin{equation}
I_{2}=\int_{r=0}^{\infty }rJ_{1}\left( ar\right) J_{1}\left( br\right)
=\delta \left( a-b\right) \frac{1}{a}\text{.}  \label{I2}
\end{equation}%
This has been generalized by Auluck\cite{Auluck} to three functions,
\begin{eqnarray}
I_{3} &=&\int_{r=0}^{\infty }rJ_{1}\left( ar\right) J_{1}\left( br\right)
J_{1}\left( cr\right)  \label{I3} \\
&=&\frac{\pi }{4c^{2}}\sin \phi P_{1}^{-1}\left( \cos \phi \right) =\frac{%
\pi }{4c^{2}}\sin ^{2}\phi \text{.}  \notag
\end{eqnarray}%
Here $c<a+b,a<b+c,b<a+c$ and $\phi $ is the angle between sides $a$ and $b$
of the triangle formed by $a,b,c$,
\begin{equation}
\sin ^{2}\phi =\frac{\left( c^{2}-\left( a-b\right) ^{2}\right) \left(
\left( a+b\right) ^{2}-c^{2}\right) }{4a^{2}b^{2}}\text{,}  \label{angle}
\end{equation}%
and $P$ is Legendre spherical harmonic. Consequently%
\begin{equation}
I_{3}=\pi \frac{\left( c^{2}-\left( a-b\right) ^{2}\right) \left( \left(
a+b\right) ^{2}-c^{2}\right) }{16a^{2}b^{2}c^{2}}\text{.}  \label{I3final}
\end{equation}

\end{document}